\documentclass[twocolumn,
aps,nofootinbib,showpacs,showkeys,preprint
tightenlines
] {revtex4-1}

\usepackage{epsf,epsfig,subfigure,graphicx,amsmath,amssymb}
\usepackage{color}
\usepackage{float}
\newcommand{\dis}[1]{\begin{equation}\begin{split}#1\end{split}\end{equation}}

\newcommand{\ie}{{\it i.e.}\ }

\begin{document}

\title{\large\bf  An exact CKM matrix related to the approximate Wolfenstein form}

\author{Jihn E. Kim$^{a,b}$ and Min-Seok Seo$^a$
\email{jekim@ctp.snu.ac.kr}
}
\affiliation{
$^a$Department of Physics and Astronomy and Center for Theoretical Physics, Seoul National University, Seoul 151-747, Korea\\
$^b$GIST College, Gwangju Institute of Science and Technology,  Gwangju 500-712, Korea \\
 }
\begin{abstract}
Noting the hierarchy between three mixing angles, $\theta_{2,3}={\cal O}(\theta_1^2)$, we present an exact form of the quark mixing matrix, replacing Wolfenstein's approximate form.  In addition, we suggest to rotate the unitarity triangle, using the weak CP phase convention where the phase is located at the (31) element $\sin\theta_1\sin\theta_2  e^{i\delta}$ while the (13) element $\sin\theta_1\sin\theta_3$ is real. For the $(ab)$ unitarity triangle, the base line ($x$-axis) is defined from the product of the first row elements, $V_{1a}V^*_{1b}$, and the angle between two sides at the origin is defined to be the phase $\delta$. 
This is a useful definition since every Jarlskog triangle has the angle $\delta$ at the origin, defined directly from the unitarity condition. It is argued that $\delta$ represents the barometer of the weak CP violation, which can be used to relate it to possible Yukawa textures.

\end{abstract}


\maketitle

\section{Introduction}\label{sec:Introduction}

The 50 years old quark mixing \cite{Cabibbo63} is the key for strange particles to decay and for the weak CP violation to be realized in the standard model (SM) \cite{CKM73,Maiani76,ChauKeung84}. Existence of CP violation is one key element for baryon number generation, the baryogenesis. At present, however, we cannot relate the weak CP violation, satisfying `baryon number conservation', to the CP violation needed for the baryogenesis employing  `baryon number violation'. Nevertheless, the weak CP violation study might hints physics far above the electroweak scale and hopefully to the CP phase appearing in the baryogenesis. This may come by observing a deviation from the SM prediction or completion of the weak CP violation in an ultraviolet completed theory beyond the SM. In any case, completing the SM is achieved by the complete determination of the quark and lepton mixing matrices, and hence determining the mixing matrices in terms of an exact form is of fundamental importance.

The original Kobayashi-Maskawa (KM) model for weak CP violation in the SM is represented by the KM unitary matrix $V_{\rm CKM}$.\footnote{We use `CKM' to represent any unitary quark-mixing matrix and `KM' for the Kobayashi-Maskawa form of Ref. \cite{CKM73}.}
The Cabibbo angle $\sin\theta_C$, the $(us)$ element of $V_{\rm CKM}$, is known to be small but not too small, $\sim 0.23$, and hence can be a good expansion parameter. A decade after the KM matrix, Wolfenstein expressed the matrix elements in terms of powers of $\lambda\simeq \sin\theta_C$ \cite{Wol83}. The Wolfenstein parametrization introduces four parameters as required: three real parameters $\lambda, A, \rho$, and an imaginary  parameter $i\eta$,
\dis{
&V_{\rm Wolf}=\left(\begin{array}{ccc}  1-\lambda^2/2 ,      &\lambda ,   &  A\lambda^3(\rho-i\eta) \\
      - \lambda, & 1-\lambda^2/2  ,  &     A\lambda^2 \\
       A\lambda^3(1-\rho-i\eta),  &  -A \lambda^2   , &1
\end{array}\right)\\
&\hskip 3cm +{\cal O}(\lambda^4).\label{eq:Wolfenstein}
}
which is very useful in fitting the weak CP violation data \cite{PData10}. Even though the imaginary number $i\eta$ is introduced, it is not a phase. If any parametrization is introducing an invariant phase, a unitary transformation should not change its determinant. The determinant of the original KM matrix has a phase \cite{CKM73}, while the Chau-Keung-Maiani form does not have a phase \cite{Maiani76,ChauKeung84}. If the phase is present, it is related to the phase of the quark mass matrix, Arg.Det.$M_q$. This can be rotated away if there exists the Peccei-Quinn symmetry \cite{PQ77}. If there is no Peccei-Quinn symmetry, one may resort to calculable vacuum angle ($\theta_{\rm weak}$) models \cite{KimRMP10} where one starts with Arg.Det.$M_q=0$. Therefore, the determinant of a well-defined quark mixing matrix better does not have a phase.

A posteriori,  the (13) and (31) elements are known to be very small ${\cal O}(\lambda^3)$. Also, it is known that the third generation is needed to have the weak CP violation. In addition, specifically, if either the (13) or (31) element is zero, there is no CP violation. In general, if any one element among nine elements of the CKM matrix is zero, then there is no weak CP  violation since one can find an appropriate phase redefinition such that all the CKM elements become real. In the Wolfenstein form, the product of (13), (22) and (31) elements, appearing among six terms of the determinant, is a barometer of CP violation,
\dis{
A^2\lambda^6\sqrt{\rho^2+\eta^2}\sqrt{(1-\rho)^2+\eta^2}e^{-i(\delta_b+\delta_t)}
\label{eq:WolfDetphase}
}
where $\tan\delta_b=\eta/\rho$ and  $\tan\delta_t=\eta/(1-\rho)$. But in the Wolfenstein form, the case (13) or (31) element vanishing is not parametrized by one parameter. Therefore, there is a need to parametrize the weak CP violation in a better form where vanishing of one parameter makes either (13) or (31) element vanishing. In fact, Qin and Ma (QM) realizes this scheme which is identical to the Wolfenstein form with the QM phase interpreted by those of Eq. (\ref{eq:WolfDetphase}), $\delta_{\rm QM}=\delta_b+\delta_t$ \cite{QinMaPLB11}.
In these $\lambda$ expanded forms, the weak CP violation occurs at order $\lambda^6$.

\section{Invariant CP phase}

\begin{figure}
\includegraphics[width=6.cm]{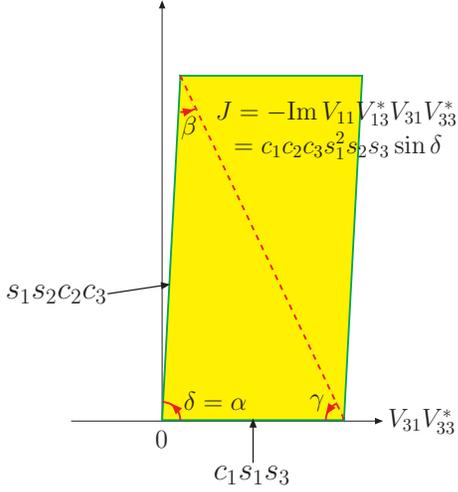} 
\caption{The unitarity triangle for $(V_{\rm KS}^\dagger V_{\rm KS})_{31}$.
} \label{fig:Utriangle}
\end{figure}

When we calculate the determinant of the approximate quark mixing matrix, the CP phase appears at order $\lambda^6$. Therefore, to keep track of the CP violation under different definitions of $V_{\rm CKM}$, we must consider the phases appearing in the higher orders as denoted below. Let us consider a modified Wolfenstein form so that the magnitudes of (13) and (31) components are represented by $\lambda^3\kappa_b$ and $\lambda^3\kappa_t$. By one parameter, $\kappa_b=0$ or $\kappa_t=0$, the (13) or (31) element becomes zero. Since vanishing of (13) or (31) element leads to no weak CP violation, it is a useful representation. In addition, if all the elements of $V_{\rm CKM}$ are real, there is no weak CP violation.  Therefore, for the weak CP violation, taking an imaginary part is essential.
The test is provided by the six contributions in the determinant of $V_{\rm CKM}$. Obviously, if all six terms in the determinant are real, there is no CP violation. Even though the whole determinant is real, the complex individual parts describe the existence of CP violation. A similar trend is observed in the decay of a particle: due to the CPT theorem the total decay rate and mass of the antiparticle is the same as those of the particle, but the individual decay rates can be different for a particle and its antiparticle.
Thus, one component among six of the determinant entries, \ie $V_{31}V_{22}V_{13}$ is the barometer, as proved below. So, it is convenient to put the phase at one place among the (13) and (31) elements. We suggest to put the phase in the (31) element, $\lambda^3\kappa_te^{i\delta}$. Then, the weak CP violation is absent if $\kappa_b=0,$ or $\kappa_t=0$, or $\delta=0$, as anticipated. In this form, the Jarlskog determinant is $\kappa_b\kappa_t\sin\delta$.  Therefore, we suggest a modified form from the Wolfenstein and QM parametrizations, valid up to ${\cal O}(\lambda^6)$,
\begin{widetext}
\dis{
\left(\begin{array}{lll}  1-\frac{\lambda^2}{2}-\frac{\lambda^4}{8}-\frac{\lambda^6}{16}(1+8\kappa_b^2), \quad &\lambda  ,   &  \lambda^3 \kappa_b\left(1+\frac{\lambda^2}{3}\right) \\ [1em]
-\lambda+\frac{\lambda^5}{2}(\kappa_t^2-\kappa_b^2) ,\quad
 & \begin{array}{l}
1-\frac{\lambda^2}{2}-\frac{\lambda^4}{8}-\frac{\lambda^6}{16} \\[0.2em]
-\frac{\lambda^4}{2}(\kappa_t^2+\kappa_b^2-2\kappa_b\kappa_t e^{-i\delta})\\[0.2em]
-\frac{\lambda^6}{12}\left(7 \kappa_b^2+\kappa_t^2-8\kappa_t \kappa_b e^{-i \delta}\right)
\end{array},      &
\begin{array}{l}
\lambda^2\left(\kappa_b-\kappa_t e^{-i\delta} \right) \\[0.2em]
 -\frac{\lambda^4}{6}(2\kappa_t e^{-i \delta}+\kappa_b)
\end{array}\\ [2.5em]
      -\lambda^3 \kappa_t e^{i\delta}\left(1+\frac{\lambda^2}{3}\right) ,  &
\begin{array}{l}
 -\lambda^2\left(\kappa_b-\kappa_t e^{i\delta} \right)\\[0.2em]
 -\frac{\lambda^4}{6}(2\kappa_b+ \kappa_te^{i\delta})
\end{array}
 , &
\begin{array}{c}
1-\frac{\lambda^4}{2}(\kappa_t^2+\kappa_b^2
      -2\kappa_b\kappa_t e^{i\delta})\\[0.2em]
-\frac{\lambda^6}{6} \left(2[\kappa_b^2+\kappa_t^2]-\kappa_t \kappa_b e^{i \delta}
\right)
\end{array}
\end{array}\right) \label{eq:KSrotated}
}
\end{widetext}
Note that our $\lambda$ is the whole element of $V_{12}$. From the absolute values of (11), (12), (13), and (31) entries ($|V_{ud}|, |V_{us}|, |V_{ub}|, |V_{td}|$) and the ratio $|V_{ub}/V_{td}|$ of the measured CKM matrix \cite{PData10}, three real parameters are determined by the quadrature,
\dis{
&\lambda=0.22527\pm 0.00092,\\
&\kappa_t=0.7349\pm 0.0141,~~\kappa_b=0.3833\pm 0.0388, \\
&\delta=89.0^{\rm o}\pm 4.4^{\rm o}
}
where we also included the phase determined by the shape of the (13) Jarlskog triangle \cite{PData10}.\footnote{In this paper, `Jarlskog triangle' is used for ${\cal O}(\lambda^6)$ triangle of Fig. \ref{fig:Utriangle} while `Jarlskog determinant' $J_{\rm Jkg}$ for that not including $\lambda^6$.}

Here, we note that row 1 or column 1 has a hierarchy of 1, $\lambda$, and $\lambda^3$. These boundary row and column are parametrized by $\theta_3$ and $\theta_2$, respectively. Therefore, assuming the hierarchy of angles, $\theta_{2,3}={\cal O}(\theta_1^2)$, the approximate unitary matrix (\ref{eq:KSrotated}) is converted to an exact unitary matrix,
\begin{widetext}
\dis{
V_{\rm KS}=\left(\begin{array}{ccc} c_1 & ~s_1c_3 & ~s_1s_3 \\ [0.2em]
 -c_2s_1 & ~e^{-i\delta}s_2s_3 +c_1c_2c_3  & ~~~ -e^{-i\delta} s_2c_3+c_1c_2s_3 \\[0.2em]
-e^{i\delta} s_1s_2  & ~-c_2s_3 +c_1s_2c_3 e^{i\delta} & ~~~c_2c_3 +c_1s_2s_3 e^{i\delta}
\end{array}\right) \label{eq:KSexact}
}
\end{widetext}
where $s_i=\sin\theta_i$ and $c_i=\cos\theta_i$. The approximate parameters $\lambda^2\kappa_t$ and $\lambda^2\kappa_b$ turn to exact parameters $s_2$ and $s_3$, respectively. We determine $\theta_1$ from $|V_{ud}|$, $\theta_2$ from $|V_{us}|$ and $|V_{ub}|$, and $\theta_3$ from $|V_{td}|$ and 
$|V_{td}/V_{ts}|$,
\dis{
&\theta_1=13.0305^{\rm o}\pm 0.0123^{\rm o}=0.227426 \pm 2.14 \times10^{-4},\\
&\theta_2=2.42338^{\rm o}\pm 0.1705^{\rm o}=0.042296 \pm 2.976 \times10^{-3},\\
&\theta_3=1.54295^{\rm o}\pm 0.1327^{\rm o}=0.027567 \pm 2.315 \times10^{-3}, \\
&\delta=89.0^{\rm o}\pm 4.4^{\rm o}.
}
Let us now scrutinize the determinant. The determinant is real, but its six elements are complex,
\dis{
V_{11}V_{22}V_{33} &=c_1^2 c_2^2c_3^2+c_1^2 s_2^2s_3^2+2c_1c_2c_3s_2 s_3 \cos{\delta}\\
 &\quad \quad\quad \quad  -c_1c_2c_3 s_1^2s_2 s_3 e^{i\delta}\\
-V_{11}V_{23}V_{32} &=c_1^2 c_2^2s_3^2+c_1^2 s_2^2c_3^2-2c_1c_2c_3s_2 s_3 \cos{\delta}\\
&\quad \quad \quad \quad +c_1c_2c_3 s_1^2s_2 s_3 e^{i\delta}\\
V_{12}V_{23}V_{31} &=s_1^2 s_2^2c_3^2-c_1c_2c_3 s_1^2s_2 s_3 e^{i\delta}\\
-V_{12}V_{21}V_{33} &=s_1^2 c_2^2c_3^2+c_1c_2c_3 s_1^2s_2 s_3 e^{i\delta}\\
V_{13}V_{21}V_{32} &=s_1^2 c_2^2s_3^2-c_1c_2c_3 s_1^2s_2 s_3 e^{i\delta}\\
-V_{13}V_{22}V_{31} &=s_1^2 s_2^2s_3^2+c_1c_2c_3 s_1^2s_2 s_3 e^{i\delta}\,. \label{eq:elementDet}
}

\noindent
If every element is real, there is no CP violating weak processes. Note that
each element of Eq. (\ref{eq:elementDet}) has an imaginary component $\pm c_1c_2c_3 s_1^2s_2 s_3 \sin\delta$ which is the real barometer of the weak CP violation. The Jarlskog triangle of Fig. \ref{fig:Utriangle} is obtained by the following entries of Eq. (\ref{eq:KSexact})
\dis{
V_{\rm KS}=\left(\begin{array}{ccc} c_1 & \times & ~s_1s_3 \\ [0.2em]
  \times  &  \times  &  \times \\[0.2em]
 -e^{i\delta}s_1s_2  &  \times  & ~c_2c_3+c_1s_2s_3 e^{i\delta}
\end{array}\right) \label{eq:KStri}
}
which gives the imaginary part of $-V_{11}V_{13}^*V_{31}V_{33}^*$ as the area of the parallelogram $J=c_1c_2c_3s_1^2s_2s_3\sin\delta$, as shown in Fig. \ref{fig:Utriangle}. If there is no imaginary part, there is no weak CP violation. Figure \ref{fig:Utriangle}, showing this imaginary part,  is  twice the $(db)$ unitarity triangle and its area represents the nature of weak CP violation. It is ${\cal O}(\lambda^6)$ times the Jarlskog determinant \cite{Jarlskog85}. The $(13)=(\bar ub)$ element of $V_{\rm KS}$ is the base line, due to $V_{11}\simeq 1$, of the unitarity triangle, and  $(31)=(\bar td)$  element, due to $V_{33}\simeq 1$,  defines our CP phase $\delta$. In this sense, the magnitude of the phase $\delta$ defines the significance of CP violation in the whole theory, \ie the significance of CP violation in the SM.

\begin{figure}
\includegraphics[width=3.cm]{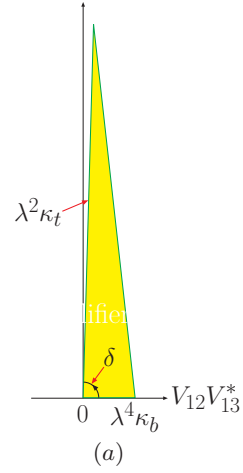}\\ \hskip -0.5cm ($a$)
\vskip 0.5cm
\includegraphics[width=8.cm]{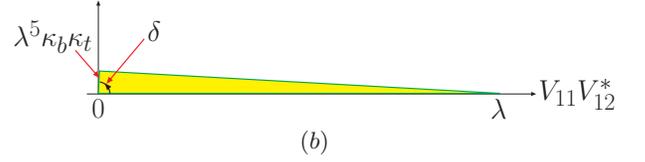}\\ ($b$)
\caption{The unitarity triangles for ($a$) $(V_{\rm KS}^\dagger V_{\rm KS})_{32}$  and
 ($b$) $(V_{\rm KS}^\dagger V_{\rm KS})_{21}$.
} \label{fig:Utriangle2}
\end{figure}

We can show that the same area $J$ also result from the $(sb), (ds), (ut), (ct),$ and $(uc)$ unitarity triangles \cite{Jarlskog85,CPbooks}. For the $(sb)$ and $(ds)$ triangles, they are shown with the approximate form, Eq. (\ref{eq:KSrotated}), in Fig. \ref{fig:Utriangle2}. As shown in Figs. \ref{fig:Utriangle} and \ref{fig:Utriangle2}, we suggest to rotate the conventional unitarity triangle so that the base line is real, \ie the (13) element.

The weak CP violation is proportional to the length squared $\lambda^6\kappa_b\kappa_t$ (the product of two side lengths). For the same side lengths, the area of the parallelogram is proportional to $\sin\delta$. Therefore, $\delta$ is really a good barometer of the strength of CP violation. Now, we can convincingly state the maximality of weak CP violation since the combined fit shows $\delta\simeq\frac{\pi}{2}$ \cite{PData10}.

\section{Yukawa textures}
The maximality of CP violation can be related to the Yukawa texture. Discrete symmetries have been considered for generating the CKM matrix from a theoretical point of view \cite{KimSeo10}.

The diagonal mass matrices, $M^{(u)}$ and  $M^{(d)}$ under the mass eigenstate bases $u^{\rm (mass)}=(u,c,t)^T$ and $d^{\rm (mass)}=(d,s,b)^T$, are
\dis{
 \frac{M^{(u)}}{m_t}=\left(\begin{array}{ccc}  \lambda^7 u   &0 ,   &  0 \\
      0 &    \lambda^4 c& 0 \\
       0 &  0&1
\end{array}\right), \
\frac{M^{(d)}}{m_b}=\left(\begin{array}{ccc}  \lambda^4 d   &0 ,   &  0 \\
      0 &    \lambda^2 s& 0 \\
       0 &  0&1
\end{array}\right) \label{eq:MuMd} }
where $u,c,d$ and $s$ are four real parameters of ${\cal O}(1)$. This mass matrices are obtained from the weak eigenstate mass matrices $\tilde M^{(u)}$ and $\tilde M^{(d)}$ by bi-unitary transformations,
\dis{
\tilde M^{(u)}=  R^{(u)\dagger}  M^{(u)}L^{(u)}  ,\quad \tilde M^{(d)}=  R^{(d)\dagger} M^{(d)} L^{(d)}
}
where $R^{(u),(d)}$ and $L^{(u),(d)}$ are unitary matrices used for the R-handed and L-handed quark fields. Our mixing matrix is given by
\dis{
V_{\rm KS} = L^{(u)} L^{(d)\dagger}.\label{eq:VLform}
}

Jarlskog achieved subtracting the real part of entries of Eq. (\ref{eq:elementDet}) by defining a commutator of mass matrices \cite{Jarlskog85},
\dis{
&C =-i[L^{(u)\dagger}M^{(u)} L^{(u)}, L^{(d)\dagger}M^{(d)} L^{(d)}],\\
&{\rm Det.}\, C= i(e^{i\delta}-e^{-i\delta})c\,s\, \kappa_t\kappa_b\,\lambda^{12}.
}
Then, the Jarlskog determinant is 
\dis{
J_{\rm Jkg}=\frac{-{\rm Det.}\,C}{2F_cF_s} \label{eq:JDet}
}
where
\dis{
& F_c= (m_t-m_c)(m_t-m_u)(m_c-m_u)/m_t^3\simeq c\lambda^4\\
& F_s= (m_b-m_s)(m_b-m_d)(m_s-m_d)/m_b^3\simeq s\lambda^2.
}
From Eqs. (\ref{eq:KSrotated}), (\ref{eq:MuMd}) and (\ref{eq:JDet}), note that $J_{\rm Jkg}$ is the paralleogram area $J$ of Fig. \ref {fig:Utriangle} devided by $\lambda^6$.

The $V_{\rm CKM}$ does not depend on the choice of unitary matrices of the R-handed quarks.
So, to glimpse the defining form of the Yukawa textures, choosing two simple cases  $R^{(u),(d)}={\bf 1}$
and $R^{(u),(d)}=L^{(u),(d)}$,  we estimate the following $u$-type and $d$-type mass matrices:

\begin{widetext}
For $R={\bf 1}$,
\dis{
\tilde M^{(u)}&=\left(\begin{array}{ccc}
    u \lambda^7 ,  & 0,  &  0 \\
      -c \lambda^5, &  c\lambda^4(1+\frac{1}{6}\lambda^2), & c \kappa_t \lambda^6 \\
       -\kappa_t e^{i \delta} \lambda^3(1+\frac{1}{3}\lambda^2), &  \kappa_t e^{i \delta} \lambda^2
       (1-\frac{\lambda^2}{6}+(\kappa_b^2-\frac{41}{360})\lambda^4), & -e^{i \delta}(1-\kappa_t\frac{\lambda^4}{2}-\kappa_t^2\frac{\lambda^6}{3})
\end{array}\right)\\
 \tilde M^{(d)} &=\left(\begin{array}{ccc}
    d \lambda^4(1+\frac{2}{3}\lambda^2) ,  & 0  , &  0 \\
      0 ,&  s\lambda^2(1+\frac{\lambda^2}{3}+(\frac{8}{45}+\frac{\kappa_b^2}{2})\lambda^4), 
      & s \kappa_b e^{i \delta} \lambda^4 (1+\frac{2}{3}\lambda^2)  \\
   0, &  \kappa_b  \lambda^2 (1+\frac{\lambda^2}{3}+(\frac{8}{45}+\kappa_b^2)\lambda^4), & -e^{i \delta}(1-\kappa_b^2\frac{\lambda^4}{2}
   -\kappa_b^2\frac{\lambda^6}{3})
\end{array}\right).\label{eq:TextureR1}
}

For $R=L$,
\dis{
\tilde M^{(u)}&=\left(\begin{array}{ccc}  (c+\kappa_t^2 \lambda) \lambda^6 ,  & -(c+\kappa_t^2 ) \lambda^5 , 
 &  \kappa_t\lambda^3 (1+\frac{1}{3}\lambda^2) \\
  -(c+\kappa_t^2 ) \lambda^5 ,&  c\lambda^4(1- \frac{1}{3} \lambda^2), & -\kappa_t \lambda^2+\frac{\kappa_t}{6}\lambda^4 +O(\lambda^6) \\
       \kappa_t\lambda^3 (1+\frac{1}{3}\lambda^2), &-\kappa_t \lambda^2+\frac{\kappa_t}{6}\lambda^4 +O(\lambda^6) ,
       & 1-\kappa_t^2\frac{\lambda^4}{2}-\kappa_t^2\frac{\lambda^6}{3}
\end{array}\right)\\  
  \tilde M^{(d)}&=\left(\begin{array}{ccc}  d \lambda^4(1+\frac{2}{3}\lambda^2),   & 0  , &  0 \\
      0 ,&  s\lambda^2+(\kappa_b+\frac{s}{3})\lambda^4+(\frac{8}{45}s+\frac{2\kappa_b^2}{3})\lambda^6,
       &  \kappa_b e^{i \delta} (-\lambda^2+(s-\frac{1}{3})\lambda^4) +O(\lambda^6)  \\
       0, &  \kappa_b e^{-i \delta} (-\lambda^2+(s-\frac{1}{3})\lambda^4) +O(\lambda^6),
       & 1-\kappa_b^2\lambda^4+\kappa_b^2(s-\frac{2}{3})\lambda^6
\end{array}\right).             \label{eq:TextureRL}
}
\end{widetext}
Equation (\ref{eq:TextureR1}) has six zeros and Eq. (\ref{eq:TextureRL}) has four zeros. These may be helpful in finding the underlying symmetries of the Yukawa texture. However, the existing textures \cite{RRR93} do not include the forms of Eqs. (\ref{eq:TextureR1}) and (\ref{eq:TextureRL}).

\section{Conclusion}
In conclusion, we obtained a very useful exact form for the quark mixing matrix which can replace the Wolfenstein form in fitting the flavor physics data. We also suggest to rotate the unitarity triangle so that two sides and the angle between them are read directly from the mixing matrix. This exact 
form can lead to a clue to the Yukawa texture of the SM at the fundamental level and the origin of CP violation.

\acknowledgments{This work is supported in part by the National Research Foundation  (NRF) grant funded by the Korean Government (MEST) (No. 2005-0093841).}

\vskip 0.5cm


\begin{thebibliography}{99}

\def\prp#1#2#3{Phys.\ Rep.\ {\bf #1} (#3) #2}
\def\rmp#1#2#3{Rev. Mod. Phys.\ {\bf #1} (#3) #2}
\def\anrnp#1#2#3{Annu. Rev. Nucl.
Part. Sci.\ {\bf #1} (#3) #2}
\def\npb#1#2#3{Nucl.\ Phys.\ {\bf B#1} (#3) #2}
\def\plb#1#2#3{Phys.\ Lett.\ {\bf B#1} (#3) #2}
\def\prd#1#2#3{Phys.\ Rev.\ {\bf D#1}, #2 (#3)}
\def\prl#1#2#3{Phys.\ Rev.\ Lett.\ {\bf #1} (#3) #2}
\def\jhep#1#2#3{J. High Energy Phys.\ {\bf #1} (#3) #2}
\def\jcap#1#2#3{J. Cos. Astropart. Phys.\ {\bf #1} (#3) #2}
\def\zp#1#2#3{Z.\ Phys.\ {\bf #1} (#3) #2}
\def\epjc#1#2#3{Euro. Phys. J.\ {\bf #1} (#3) #2}
\def\ijmp#1#2#3{Int.\ J.\ Mod.\ Phys.\ {\bf #1} (#3) #2}
\def\mpl#1#2#3{Mod.\ Phys.\ Lett.\ {\bf #1} (#3) #2}
\def\apj#1#2#3{Astrophys.\ J.\ {\bf #1} (#3) #2}
\def\nat#1#2#3{Nature\ {\bf #1} (#3) #2}
\def\sjnp#1#2#3{Sov.\ J.\ Nucl.\ Phys.\ {\bf #1} (#3) #2}
\def\apj#1#2#3{Astrophys.\ J.\ {\bf #1} (#3) #2}
\def\ijmp#1#2#3{Int.\ J.\ Mod.\ Phys.\ {\bf #1} (#3) #2}
\def\apph#1#2#3{Astropart.\ Phys.\ {\bf B#1}, #2 (#3)}

\bibitem{Cabibbo63} N. Cabibbo, \prl{10}{531}{1963}.  It can be glimpsed from
a footnote of M. Gell-Mann and M. Levy, Nuovo Cim. {\bf 16} (1961) 705.

\bibitem{CKM73} M. Kobayashi and T. Maskawa, Prog. Theor. Phys. {\bf 49}, 652 (1973).

\bibitem{Maiani76} L. Maiani, in {\it Proceedings of the 1977 International Symposium on Lepton and Photon Interactions at High Energies} (DESY, Hamburgh, 1977), p. 867; \plb{62}{183}{1976}.

\bibitem{ChauKeung84} L. L. Chau and W. Y. Keung, \prl{53}{1802}{1984}.

\bibitem{Wol83} L. Wolfenstein, \prl{51}{1945}{1983}.

\bibitem{PData10} K. Nakamura {\it et al.} (Particle Data Group),  J. Phys. G 37, 075021 (2010).

\bibitem{PQ77} R. D. Peccei and H. R. Quinn, \prl{38}{1440}{1977}.

\bibitem{KimRMP10} J. E. Kim, \prp{150}{1}{1987}. For a recent review, see, J. E. Kim and G. Carosi, \rmp{82}{557}{2010} [arXiv: 0807.3125[hep-ph]], and references therein.

\bibitem{QinMaPLB11} N. Qin, B. Q. Ma, \plb{695}{194}{2011} [arXiv:1011.6412 [hep-ph]].

\bibitem{Jarlskog85} C. Jarlskog, \prl{55}{1039}{1985}.

\bibitem{CPbooks} G. C. Branco, L. Lavoura and J. P. Silva, {\it CP Violation},  Int. Ser. Monogr. Phys.  {\bf 103} (1999) 1--536; I. I. Bigi and A. I. Sanda, 
{\it CP violation}, Cambridge Monographs on Particle Physics and Cosmology (2009) 1-506.


\bibitem{KimSeo10} J. E. Kim and M.-S. Seo, \jhep{1102}{097}{2011} [arXiv:1005.4684], and references therein.

\bibitem{RRR93} P. Ramond, R. G. Roberts, and G. G. Ross, \npb{406}{19}{1993} [hep-ph/9303320].





\end{thebibliography}
\end{document}